\documentclass[aps,nofootinbib,prd,pacs]{revtex4} 
\usepackage{amsfonts}
\usepackage{amsmath}
\usepackage{bbm}
\usepackage{graphicx}
\usepackage{pstricks}
\usepackage{epsfig}

\newcommand{\be}{\begin{equation}}
\newcommand{\ee}{\end{equation}}
\newcommand{\beq}{\begin{eqnarray}}
\newcommand{\eeq}{\end{eqnarray}}

\def\lsim{\hbox{ \raise.35ex\rlap{$<$}\lower.6ex\hbox{$\sim$}\ }}
\def\gsim{\hbox{ \raise.35ex\rlap{$>$}\lower.6ex\hbox{$\sim$}\ }} 

\begin{document}
\title{Unstable Anisotropic Loop Quantum Cosmology} \author{William
  Nelson\footnote{William.Nelson@kcl.ac.uk} and Mairi
  Sakellariadou\footnote{Mairi.Sakellariadou@kcl.ac.uk}}
\affiliation{King's College London, Department of Physics, Strand WC2R
  2LS, London, U.K.}

\begin{abstract}
\vspace{.2cm}
\noindent
We study stability conditions of the full Hamiltonian constraint
equation describing the quantum dynamics of the diagonal Bianchi~I
model in the context of LQC. Our analysis has shown robust evidence of
an instability in the explicit implementation of the difference
equation, implying important consequences for the correspondence
between the full LQG theory and LQC. As a result, one may question the
choice of the quantisation approach, the model of lattice refinement,
and/or the r\^ole of the ambiguity parameters; all these should in
principle be dictated by the full LQG theory.
\end{abstract}

\maketitle

\section{Introduction}
Loop Quantum Gravity (LQG)~\cite{rovelli2004} is a non-perturbative,
background independent, canonical quantisation of General Relativity
in four space-time dimensions. Even though the full theory of LQG is
not yet complete, its successes encourage the application of LQG
techniques to mini-superspaces obtained by a symmetry reduction.  The
application of LQG to the cosmological sector is known as Loop Quantum
Cosmology (LQC)~\cite{Ashtekar:2003hd,Bojowald:2002gz}.  In the
homogeneous and isotropic cosmological models with a massless scalar
field, which plays the r\^ole of an internal time parameter according
which other physical quantities ``evolve'', quantum geometry effects
of the full LQG theory lead to a repulsive force in the Planckian
regime. Thus, the big bang singularity is resolved and replaced by a
quantum bounce~\cite{Ashtekar:2006wn}. The underlying discreteness of
LQC is the key element for the existence of the quantum bounce;
similar results have thus been also obtained in the context of other
models.

LQC quantum dynamics are determined by a difference, rather than a
differential, equation, as a result of quantum geometry
effects. However, such effects can be neglected as one departs from
the Planckian regime, and quantum dynamics can then be well
approximated by the Wheeler-DeWitt (WDW) differential equation.  LQC
is formulated in terms of SU(2) holonomies of the connection and
triads. In the ``old'' quantisation, the quantised holonomies were
taken to be shift operators {\it with a fixed magnitude}, but later it
was found that this leads to problematic instabilities in the
continuum semi-classical limit, where the WDW wave-function becomes a
good approximation to the difference equation of LQC.  In a dynamical
equation closer to what is expected to be obtained from the full LQG
theory, lattice refinement would take place during the evolution,
since full Hamiltonian constraint operators generally create new
vertices of a lattice state in addition to changing their edge labels.
The effect of the refinement of the discrete lattice has been modelled
and the elimination of the instabilities in the continuum era has been
explicitly shown~\cite{Nelson:2007um,Nelson:2007wj,Nelson:2008vz}.
Lattice refinement leads to new dynamical difference equations which,
in general, do not have a uniform step-size making their study quite
involved.  In contrast to isotropic models, which can be understood in
terms of wave-functions on a one-dimensional discrete mini-superspace,
anisotropic models with higher-dimensional mini-superspaces, can be
more subtle.  For the partial difference equations of anisotropic
models, stability issues can turn out to be more serious than in
isotropic ones, leading to consistency tests, and thus restricting
possible quantisation freedom.  In Ref.~\cite{Nelson:2008bx} we have
proposed a numerical method, based on Taylor expansions, which
provides the necessary information to calculate the wave-function at
any given lattice point.  We have developed~\cite{Nelson:2008bx}
numerical schemes for both the one-dimensional homogeneous and isotropic
cosmological case, which has analytic solutions, as well as the
two-dimensional case of a Schwarszchild interior, which cannot be
exactly solved.

LQC issues of the Bianchi type I models, the simplest among
anisotropic cosmologies, have been also investigated. Besides their
simplicity, such models are very interesting for addressing the issue
of space-like singularities in the context of the full LQG theory.  As
in the isotropic case, a massless scalar field plays the r\^ole of an
internal time parameter. Recent analysis~\cite{Ashtekar:2009vc} has
shown that the big bang singularity is solved by quantum gravity
effects, while LQC dynamics is well approximated by that of the WDW
theory once quantum geometry effects become negligible.

The aim of this paper is to analyse the stability conditions of the
solutions to the full Hamiltonian constraint. Unstable ({\sl i.e.},
growing) solutions would indicate unphysical spurious solutions, for
which there is no correspondence between the difference (valid in the
LQC regime) and the differential (WDW) equations. This would indicate
an inconsistency between the full LQG theory and the mini-superspace
LQC approach, implying the possibility of a weakness of the employed
quantisation approach.  This work is organised as follows:  In
Section~II we outline the basic formalism of LQG and LQC. In
Section~III we perform a stability analysis. We summarise our results
and we discuss the outcome of our findings in Section~IV.

\section{Basics of the LQG/LQC formalism}\label{sec:basics}
Let us restrict ourselves to diagonal Bianchi~I metrics, for which
space-time metric in Cartesian coordinates, $\tau, x_i$
(i=1,2,3), reads
\be
{\rm d}s^2=-N^2{\rm d}\tau^2+\sum_{i=1}^3 a_i^2{\rm d}x_i^2~,
\ee
where $N$ is the lapse function and $a_i$ (with $i=1,2,3$) stand for
the three directional scale factors. Following
Ref.~\cite{Ashtekar:2009vc} we choose $\tau$ to satisfy $\Box\tau=0$.

LQG/LQC are based on a Hamiltonian formulation of General Relativity,
with basic variables an SU(2) valued connection $A^i_a$ and the
conjugate momentum variable which is a densitised triad $E^a_i$ , a
derivative operator quantised in the full LQG theory in the form of
fluxes.  As for any quantisation scheme based on a Hamiltonian
framework or an action principle, for the homogeneous flat model one
should regularise the divergences which appear due to the homogeneity
as the action and Hamiltonian are integrated over spatial
hyper-surfaces.  We thus restrict spatial homogeneity and Hamiltonian
to an elementary cell ${\cal V}$, which we choose so that its edges
lie along the fixed coordinate axis $x_i$ (with $i=1,2,3$). In
addition, we fix a fiducial flat metric $^0q_{ab}$, with line element
\be
{\rm d}s_0^2=\sum_{i=1}^3 {\rm d}x_i^2~.
\ee
The lengths of the three edges of the elementary cell ${\cal V}$ and
its volume, as measured by the fiducial flat metric $^0q_{ab}$, are
denoted by $L_i$ (with $i=1,2,3$) and $V_0=L_1L_2L_3$, respectively.

The densitised triad carries information about the spatial geometry,
encoded in the three-metric, while the connection carries information
about the spatial curvature, in the form of the spin-connection and
the extrinsic curvature.  We introduce physical triads $^0e^a_i$ and
their dual (${^0}e^a_i \ ^0\omega^j_a=\delta^j_i$) co-triads
$^0\omega^i_a=D_ax^i$, satisfying $^0q_{ab} ={^0}\omega^i_a
\ ^0\omega^j_b\delta_{ij}$.  Note that $i$ refers to the Lie algebra
index and $a$ is a spatial index with $a,i=1,2,3$. The physical
co-triads are given by $\omega^a_i=a^i \ {^0}\omega^i_a$, and the physical
three-metric by $q_{ab}=\omega^i_a \omega^j_b \delta_{ij}$.

The six-dimensional phase space is defined through the SU(2)
connection $A^i_a$ and the triad $E^a_i$ given by
\beq
A^i_a &=& c^i(L^i)^{-1} \ {^0}\omega^i_a\nonumber\\
E^a_i&=&p_iL_iV_o^{-1}\sqrt{^0q}\ {^0}e^a_i~,
\eeq
where the connection components $c_i$ and the momenta $p_i$ are
constants; $q=(p_1p_2p_3) ^0q V_0^{-1}$ stands for the determinant of
the physical spatial metric $q_{ab}$. The three momenta $p_i$ are
related to the three scale factors through
\beq
p_1&=&{\rm sgn} (a_1)|a_2a_3|L_2L_3\nonumber\\
p_2&=&{\rm sgn} (a_2)|a_1a_3|L_1L_3\nonumber\\
p_3&=&{\rm sgn} (a_3)|a_1a_2|L_1L_2~.
\eeq
The pairs $c^i, p_i$ (with $i=1,2,3$) satisfy the Poisson brackets
relations: 
\be 
\{c^i,p_j\}=8\pi G\gamma \delta^i_j~, 
\ee 
with $\gamma$ the Barbero-Immirzi parameter.

Two of the constraints of the full LQG theory, namely the Gauss and
the diffeomorphism constraints are identically satisfied and one is
therefore left with the Hamiltonian constraint, as for the isotropic
case. Restricting the integration to the fiducial cell ${\cal V}$, the
Hamiltonian constraint reads
\be
\label{h-c}
{\cal C}=\int_{\cal V} N ({\cal H}_{\rm grav}+{\cal H}_{\rm
  matter}){\rm d}^3x~, 
\ee
where ${\cal H}_{\rm grav}$ and ${\cal H}_{\rm matter}$ stand for the
gravitational and the matter parts of the constraint densities,
respectively. The lapse function $N$ is $N=\sqrt{|p_1p_2p_3|}$.

Since Bianchi~I models are spatially flat, the matter part of the
Hamiltonian constraint can be written as~\cite{Ashtekar:2009vc}
\be
\label{ham-const}
{\cal H}_{\rm grav}=-{\sqrt{^0q}\over 8\pi
  G\gamma^2\sqrt{p_1p_2p_3}V_0}(p_1p_2c_1c_2+p_1p_3c_1c_3+p_2p_3c_2c_3)~.
\ee
The matter part of the Hamiltonian constraint is~\cite{Ashtekar:2009vc}
\be
{\cal H}_{\rm matter}=\sqrt{q}\rho_{\rm matter}~,
\ee
where $\rho_{\rm matter}$ is the matter energy density of the matter
field, chosen to be a massless scalar field $T$; 
\be
\rho_{\rm  matter}={p^2_T\over 2 |p_1p_2p_3|}~,
\ee
with $p_T$ the canonically conjugate momentum of $T$.  The scalar
field $T$ can be considered as an evolution parameter in the classical
theory, and as a viable internal time parameter in the subsequent
quantum theory.  The justification for this choice lies in the fact
that since $p_T$ is a constant of motion, $T$ grows linearly in time
$\tau$, for any solution to the field equations.
The full Hamiltonian constraint, Eq.~(\ref{ham-const}) can then be
finally written as~\cite{Ashtekar:2009vc}
\be 
{\cal H}=-{1\over 8\pi
  G\gamma^2}(p_1p_2c_1c_2+p_1p_3c_1c_3+p_2p_3c_2c_3)+{p_T^2\over 2}~.
\ee

Let us proceed with the quantum kinematics of Bianchi~I LQC. The
gravitational part of the kinematic Hilbert space, ${\cal H}_{\rm
  kin}^{\rm grav}$, can be expressed in the momentum, $p_i$ (with
$i=1,2,3$), representation. Given an orthonormal basis states
$|p_1,p_2,p_3\rangle$, which are eigenstates of quantum geometry,
consider a linear combination
\be
|\Psi\rangle=\sum_{p_1,p_2,p_3}\Psi(p_1,p_2,p_3)|p_1,p_2,p_3\rangle~,
\ee
with finite norm, namely
\be
\sum_{p_1,p_2,p_3}|\Psi(p_1,p_2,p_3)|^2<\infty~,
\ee
and
\be
\langle p_1,p_2,p_3|p'_1,p'_2,p'_3\rangle=\delta_{p_1 p_1'}\delta_{p_2
  p_2'}\delta_{p_3 p_3'}~. 
\ee

The action of the elementary operators, which are the three momenta
$p_i$ (with $i=1,2,3$) and the holonomies $h_i^{(\ell)}$ along edges
parallel to the three axis $x_i$ (with $i=1,2,3$) --- completely
determined by almost periodic ($\ell$ is any real number) functions
$\exp(i\ell c_j)$ of the connection --- is given
by~\cite{Ashtekar:2009vc}
\beq
\hat{p}_1|p_1,p_2,p_3\rangle&=&p_1|p_1,p_2,p_3\rangle\nonumber\\ 
\widehat{\exp(i\ell c_1)}|p_1,p_2,p_3\rangle&=&|p_1-8\pi
G\gamma\hbar\ell, p_2,p_3\rangle~, 
\eeq
and similarly for $\hat{p}_2, \widehat{\exp(i\ell c_2)}$ and
$\hat{p}_3, \widehat{\exp(i\ell c_3)}$.

One has then to build the quantum analogue of the Hamiltonian
constraint, along the lines of the isotropic case. To do so, one has
to find the operator on the gravitational sector of the kinematic
Hilbert space, corresponding to the curvature $F_{ab}^{\ \ k}$ of the
connection $A^i_a$, given by
\be
F_{ab}^{\ \ k}=2\partial _{[a}A_{b]}^{\ \ k}+\epsilon_{ij}^{\ \ k}A^i_aA^j_b~.
\ee

As it is known from the isotropic case, the connection operator does
not exist in LQG/LQC; we cannot take the limit of the area enclosed by
a plaquette to go to zero, since the minimum area enclosed by the
plaquette is the nonzero eigenvalue $\Delta p_{\rm Pl}^2$ (with
$\Delta$ a dimensionless number, $\Delta=4\sqrt{3}\pi\gamma$) of the
area operator.  To single out a unique plaquette of the many ones
enclosing an area $\Delta p_{\rm Pl}^2$ on each of the three faces of
the elementary cell ${\cal V}$, we will use the natural gauge fixing
available for the diagonal Bianchi~I case, and a correspondence
between kinematic states in LQG and LQC. In this way, one obtains that
the curvature operator reads~\cite{Ashtekar:2009vc}
\be
\hat{F}_{ab}^{\ \ k}=\epsilon_{ij}^{\ \ k}\left({\sin{\bar\mu c}\over\bar\mu
  L}\ ^0\omega_a\right)^i 
\left({\sin{\bar\mu c}\over\bar\mu
  L}\ ^0\omega_b\right)^j~,
\ee
where
\be
\left({\sin{\bar\mu c}\over\bar\mu
  L}\ ^0\omega_a\right)^i={\sin{\bar\mu}^ic^i\over{\bar\mu}^iL^i}\ ^0\omega^i_a~,
\ee
with
\beq
\bar\mu_1&=&\sqrt{{|p_1|\Delta l^2_{\rm Pl}\over |p_2p_3|}}~,\nonumber\\
\bar\mu_2&=&\sqrt{{|p_2|\Delta l^2_{\rm Pl}\over |p_1p_3|}}~,\nonumber\\
\bar\mu_3&=&\sqrt{{|p_3|\Delta l^2_{\rm Pl}\over |p_1p_2|}}~.
\eeq
The functional dependence of $\bar\mu_i$ on $p_i$ is essential since
otherwise quantum dynamics can depend on the choice of the fiducial
cell ${\cal V}$.

Consequently, one can now write the quantum analogue of the full
Hamiltonian constraint, Eq.~(\ref{h-c}). It
reads~\cite{Ashtekar:2009vc}
\be
-\hbar^2\partial_T^2\Psi(\vec\lambda,T)=\Theta\Psi(\vec\lambda,T)~,
\ee
where $\Theta=-{\cal C}_{\rm grav}$.

To simplify the gravitational sector of the Hamiltonian constraint,
one can introduce the volume of the elementary cell ${\cal V}$ as
one of the arguments of the wave function. Let us then
set~\cite{Ashtekar:2009vc} 
\be
\nu=2\lambda_1\lambda_2\lambda_3~,
\ee
which is directly related to the volume of ${\cal V}$, namely
\be
\hat V\Psi(\lambda_1,\lambda_2,\nu)=2\pi|\gamma|\sqrt{\Delta}|\nu|l^3_{\rm pl}
\Psi(\lambda_1,\lambda_2,\nu)~,
\ee 
with $\gamma={\rm sgn}(p_1p_2p_3)|\gamma|$.
Thus, the new configuration variables will be $\lambda_1, \lambda_2,
\nu$.

In the next section, we will write out explicitly the full Hamiltonian
constraint and we will then study the stability of its solutions.

\section{Stability analysis}
The basic difference equation arising from the loop quantisation of
the Bianchi~I model reads~\cite{Ashtekar:2009vc}
\beq\label{eq:diff_eqn}
 \partial_T^2 \Psi \left( \lambda_1, \lambda_2, \nu ; T \right) &=&
 \frac{\pi G}{2} \sqrt{\nu} \Bigl[ \left( \nu+2\right) \sqrt{\nu +4}
   \Psi^+_4 \left(\lambda_1,\lambda_2,\nu;T\right) -\left(
   \nu+2\right) \sqrt{\nu} \Psi^+_0\left(
   \lambda_1,\lambda_2,\nu;T\right) \nonumber \\ && - \left(
   \nu-2\right) \sqrt{\nu} \Psi^-_0 \left(
   \lambda_1,\lambda_2,\nu;T\right)
   +\left(\nu-2\right)\sqrt{|\nu-4|}\Psi^-_4 \left(
   \lambda_1,\lambda_2,\nu;T\right) \Bigr]~,
\eeq
where
\beq
 \Psi^+_4 \left(\lambda_1,\lambda_2,\nu;T\right) &=& \sum_{i\neq
   j=(0,1,2)} \Psi \left( a_i \lambda_1, a_j
 \lambda_2,\nu+4;T\right) \nonumber \\
 \Psi^-_4 \left(\lambda_1,\lambda_2,\nu;T\right) &=& \sum_{i\neq
   j=(-3,-2,0)} \Psi \left( a_i \lambda_1, a_j
 \lambda_2,\nu-4;T\right) \nonumber \\
 \Psi^+_0 \left(\lambda_1,\lambda_2,\nu;T\right) &=& \sum_{i\neq
   j=(-1,0,1)} \Psi \left( a_i \lambda_1, a_j \lambda_2,\nu;T\right)
 \nonumber \\
 \Psi^-_0 \left(\lambda_1,\lambda_2,\nu;T\right) &=& \sum_{i\neq
   j=(-2,0,3)} \Psi \left( a_i \lambda_1, a_j \lambda_2,\nu;T\right)~,
\eeq
and the functions $a_i$ have been defined as follows:
\beq\label{eq:a}
 a_{-3} \equiv \left( \frac{\nu-4}{\nu-2}\right)~,\ \ \ a_{-2} \equiv
 \left( \frac{\nu-2}{\nu}\right)~, \ \ \ a_{-1} \equiv \left(
 \frac{\nu}{\nu+2}\right)~, \nonumber \\ a_{0} \equiv 1~, \ \ \ a_{1}
 \equiv \left( \frac{\nu+4}{\nu+2}\right)~, \ \ \ a_{2} \equiv \left(
 \frac{\nu+2}{\nu}\right)~, \ \ \ a_{3} \equiv \left(
 \frac{\nu}{\nu-2}\right)~.
\eeq
Numerical evolution can in principle be carried out by restricting
to the positive octant ($\lambda_1\geq 0, \lambda_2\geq 0, \nu\geq 0$),
thus eliminating the ${\rm sgn}(\lambda_i)$ factors which are otherwise
appearing in various terms.

\begin{figure}
 \begin{center}
  \includegraphics[scale=1.0]{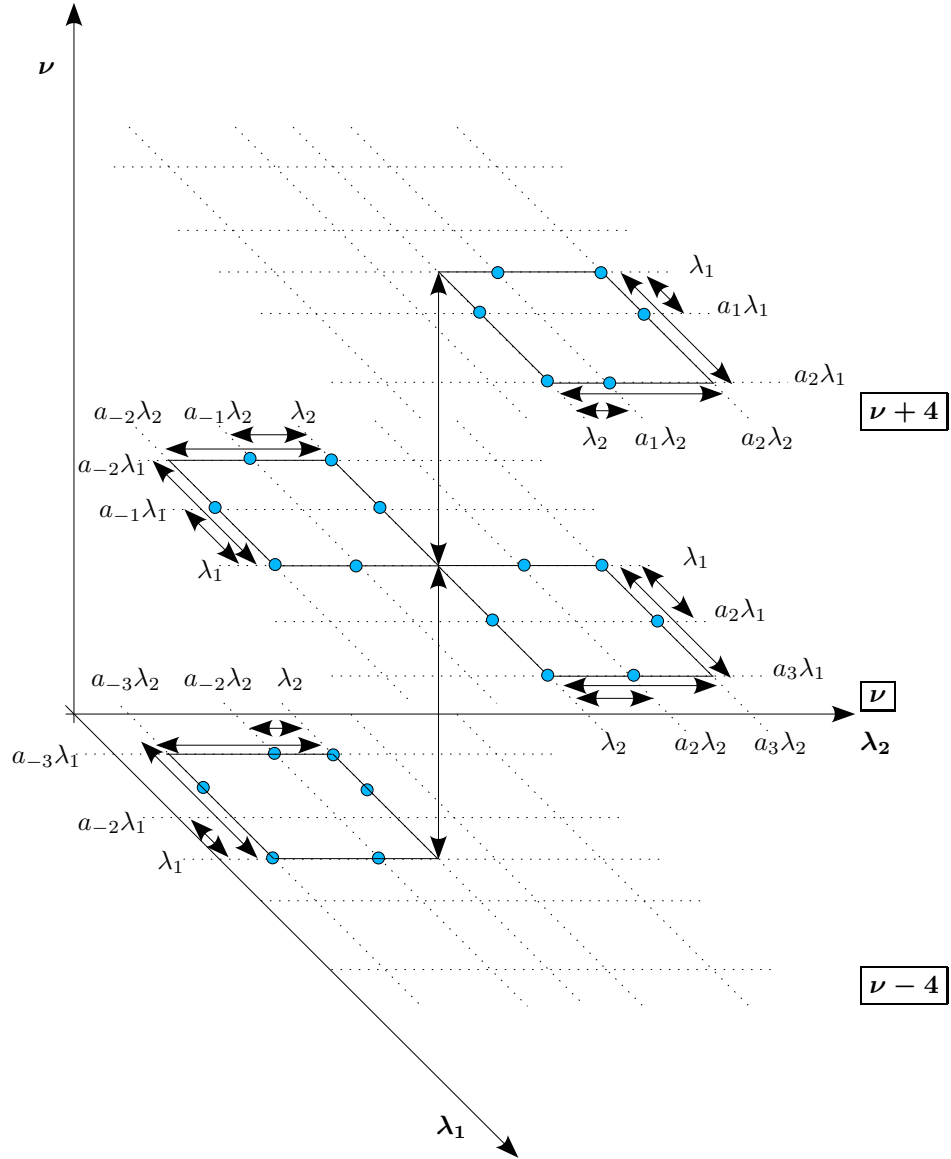}
  \caption{\label{fig1} The geometry of the points used in the
    difference equation that results from the Hamiltonian constraint,
    for the Bianchi~I model.}
 \end{center}
\end{figure}

Here we wish to examine the stability of the vacuum solutions, in
which case the solution is static, namely
$\Psi\left(\lambda_1,\lambda_2,\nu;T\right) =
\Psi\left(\lambda_1,\lambda_2,\nu\right)$, and Eq.~(\ref{eq:diff_eqn})
becomes
\be\label{eq:diff_eqn2} 
\Psi^+_4\left(\lambda_1,\lambda_2,\nu\right) =
\sqrt{ \frac{\nu}{\nu+4} }\Psi^+_0\left(\lambda_1,\lambda_2,\nu\right)
+\left(\frac{\nu-2}{\nu+2}\right) \sqrt{ \frac{\nu}{\nu+4}
}\Psi^-_0\left(\lambda_1,\lambda_2,\nu\right)
-\left(\frac{\nu-2}{\nu+2}\right) \sqrt{ \frac{|\nu-4|}{\nu+4} }
\Psi^-_4\left(\lambda_1,\lambda_2,\nu\right)~, 
\ee
for $\nu \neq 0$; otherwise the above equation must be multiplied by
$\sqrt\nu$, thus corresponding to the classical singularity.
The geometry of this difference equation is drawn in Fig.~\ref{fig1}.
Equation~(\ref{eq:diff_eqn2}) can be used to evaluate the value of the
wave-function on the $\nu+4$ plane, given suitable boundary conditions
on the $\nu$ and $\nu-4$ planes. The requirement that the arguments
must be positive ({\sl i.e.}, $\lambda_1\ge 0$, $\lambda_2\ge 0, \nu
\geq 0$) reduces the required number of boundary conditions.  For the
purpose of our work, it is sufficient to consider starting from a
plane in which $\nu-4 > 0$.

In addition to specifying the boundary conditions on the $\nu$ and
$\nu-4$ planes, we are also required to specify the value at five of
the points given in
$\Psi^+_4\left(\lambda_1,\lambda_2,\nu\right)$. There are in total
$23$ values that are required and with such initial data the
difference equation, Eq.~(\ref{eq:diff_eqn}), can be used to evaluate
the $24^{\rm th}$ point. Once this point has been evaluated, it can be
used to ``move'' the central point and evaluate the wave-function at
subsequent positions in the $\nu+4$ plane. In this way the difference
equation can be used to find the wave-function that is consistent with
the Hamiltonian constraint, Eq.~(\ref{eq:diff_eqn}), and the boundary
conditions. In principle, this procedure can be iterated to evaluate
the consistent wave-function for all subsequent $\nu$-planes, however
the stability of the difference equation can be investigated even at
this first iteration. 

\begin{figure}
 \begin{center}
  \includegraphics[scale=1.0]{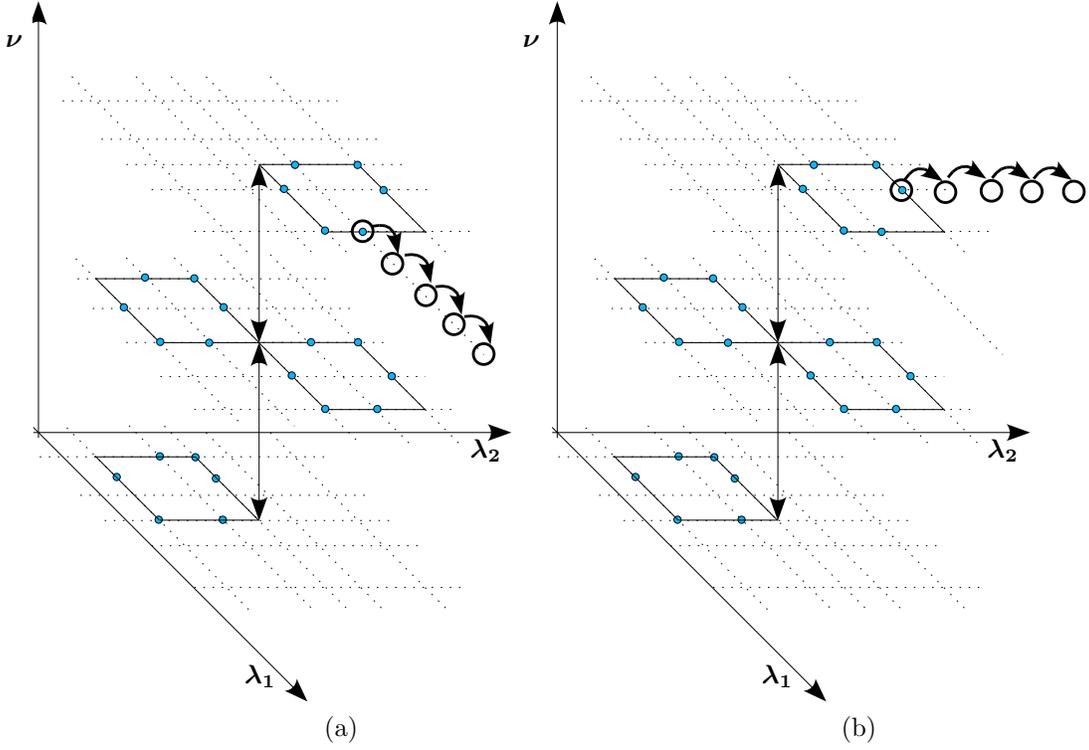}
  \caption{\label{fig2} The difference equation gives us a point in
    the $\nu+4$ plane, given the required $23$ points.  Exactly which
    point is calculated via the difference equation is somewhat
    arbitrary and essentially describes the way in which the $\nu+4$
    plane is calculated. In the l.h.s. scheme (a) the point $\left(
    a_2\lambda_1, a_1\lambda_2,\nu+4\right)$ is calculated, in which
    case the $\nu+4$ plane is evaluated first along constant
    $\lambda_2$. In the r.h.s. scheme (b) the point chosen is $\left(
    a_1\lambda_1, a_2\lambda_2,\nu+4\right)$ and the $\nu+4$ plane
    would be evaluated first along constant $\lambda_1$.}
 \end{center}
\end{figure}

As shown in Fig.~\ref{fig1}, there is a choice to be made as to which
point in the $\nu+4$ plane is to be calculated from the difference
equation. This choice amounts to deciding whether to increase
$\lambda_1$ or $\lambda_2$ first, when populating the $\nu+4$
plane. From the point of view of the plane, the difference equation,
Eq.~(\ref{eq:diff_eqn}), can be seen as progressively evaluating the
wave-function at points first along either the $\lambda_1$ direction
or the $\lambda_2$ one (see, Fig.~\ref{fig2}).  In this sense, we can
consider Eq.~(\ref{eq:diff_eqn}) as an ``evolution'' equation of a
wave-function with respect to either $\lambda_1$ or $\lambda_2$,
subject to suitable boundary conditions.  It is important to realise
however that this ``evolution'' has only to do with the order in which
the points are evaluated and is not related, in any way, to evolution
of the wave-function with respect to time.

With this view, standard von Neumann stability analysis can be
preformed on Eq.~(\ref{eq:diff_eqn}), to see if the system is
stable~\cite{Bojowald:2003dn,Rosen:2006bga}. Here however caution is
necessary.  Von Neumann's analysis is typically used to see if there
are growing mode solutions to a particular {\it discretised} version
of an underlying differential equation. In this case, the difference
equation is the fundamental evolution equation, which can be
approximated by a differential equation (the anisotropic Wheeler-DeWitt
equation~\cite{Ashtekar:2009vc}) in a suitable limit.  In standard
numerical implementations of differential equations, the stability of
the system is important only because artificial numerical rounding
errors can grow to dominate the behaviour of the solution, however the
situation here is very different. In principle, the difference
equation, Eq.~(\ref{eq:diff_eqn}), is exact and hence all solutions
should be considered, however in practise we wish to restrict only to
those solutions that closely approximate General Relativity at large
scales. This makes the use of von Neumann stability analysis useful,
since we are comparing a particular difference equation, with the
differential equation it approximates, however it is important to
remember that the motivation is very different than in standard numerical
analysis.

For homogeneous and isotropic cosmologies, a local stability analysis
of the corresponding difference equation to determine the behaviour of
spurious solutions was performed in Ref.~\cite{Vandersloot:2005kh},
using higher order spin $J$ representations of the holonomies for the
quantisation.  It was found~\cite{Vandersloot:2005kh} that the use of
higher spin holonomies to regulate the gravitational part of the
constraint operator leads to modifications, which are qualitatively
similar to those of the inverse scale factor. Stability analysis has
shown that the $J=1$ difference equation is not locally stable.  To
further determine whether these spurious solutions represent a problem
with the quantisation, the authors of Ref.~\cite{Vandersloot:2005kh}
have studied the physical inner product, since unphysical solutions
would have either vanishing or infinite physical norm and would be
modded out of the physical Hilbert space.  For the cases of Bianchi~I
locally rotationally symmetric cosmology and that of the Schwarzschild
interior geometry, a von Neumann stability analysis of a difference
equation obtained by a previous quantisation approach was carried out
in Ref.~\cite{Rosen:2006bga}, where there were identified large
regions in space-time that have generically instabilities.  In what
follows, we will look for spurious solutions to
Eq.~(\ref{eq:diff_eqn}), in the sense that they do not approximate
solutions to the relevant Wheeler-DeWitt equation in the large volume
limit.

As in standard von Neumann stability analysis, we will decompose the
solutions of the difference equation, Eq.~(\ref{eq:diff_eqn2}), into
Fourier modes and look for growing modes. Specifically, we consider
the ansatz
\be
 \Psi\left(\lambda_1,\lambda_2,\nu\right) = T\left(\lambda_1\right)
 \exp \left( i \left( \omega \lambda_2 + \chi \nu \right)\right)~, 
\ee
where we have chosen the $\lambda_1$ direction to be the direction in
which the $\nu+4$ plane is ``evolved''. Using the above ansatz,
Eq.~(\ref{eq:diff_eqn2}) becomes
\beq 
\label{dif-eq}
 e^{4\chi i} \sum_{i\neq j = \left(0,1,2\right)} T\left(a_i
\lambda_1\right) e^{ i\left( \omega a_j \lambda_2 + \chi \nu\right)}&=&
\sqrt{ \frac{\nu}{\nu + 4} } \sum_{i\neq j =\left( -1,0,1\right)}
T\left( a_i \lambda_1\right) e^{i\left(\omega a_j \lambda_2 + \chi
  \nu\right)} \nonumber \\
&& + \left(\frac{\nu-2}{\nu+2}\right) \sqrt{ \frac{\nu}{\nu + 4} }
 \sum_{i\neq j =\left( -2,0,3\right) } T\left( a_i \lambda_1\right)
 e^{i\left(\omega a_j \lambda_2 + \chi \nu\right)}\nonumber\\
&& -\left(\frac{\nu-2}{\nu+2}\right)\sqrt{ \frac{|\nu-4|}{\nu + 4} }
 e^{-4\chi i} \sum_{i\neq j =\left( -3,-2,0\right) } T\left( a_i
 \lambda_1\right) e^{i\left(\omega a_j \lambda_2 + \chi \nu\right)}~.
 \nonumber \\
\eeq
To simplify each of the summations, we proceed as follows:
\beq
\sum_{i\neq j=\left(-1,0,1\right)} T\left(
 a_i\lambda_1\right)e^{i\omega \lambda_2} &=&
 T\left(a_1\lambda_1\right) \left( e^{i\omega \lambda_2} + e^{i\omega
   a_{-1} \lambda_2}\right)\nonumber\\
&& +~T\left( a_0\lambda_1\right) \left(
 e^{i\omega a_{-1}\lambda_2} + e^{i\omega a_1 \lambda_2}\right)
 \nonumber \\ && +~T\left( a_{-1}\lambda_1\right) \left( e^{i\omega
   a_1\lambda_2} + e^{i\omega \lambda_2}\right)~,
\eeq
which becomes
\beq
\sum_{i\neq j=\left(-1,0,1\right)} T\left( a_i \lambda_1\right)
 e^{i\omega\lambda_2} &=& 2e^{i\omega \lambda_2} \Biggl[ T\left( a_1
   \lambda_1\right) e^{ -
     \frac{i\omega\lambda_2\left(a_1-1\right)}{2}} \cos \left( \frac{
     \omega \lambda_2 \left(a_1-1\right)}{2}\right) \nonumber \\ &&
\ \ \ \ \ \ \ \ \ \  
+   T\left(a_0\lambda_1\right) \cos \left( \omega \lambda_2\left(a_1 -
   1\right)\right)\nonumber\\ &&\ \ \ \ \ \ \ \ \ \ 
+ T\left(a_1\lambda_1\right) e^{\frac{i\omega
       \lambda_2\left(a_1-1)\right)}{2}} \cos \left( \frac{\omega
     \lambda_2 \left(a_1-1\right)}{2}\right) \Biggr]~,
\eeq
where we made use that
\be
a_{-3}-1 = -\left( a_3 - 1\right)~, \ \ \ a_{-2}-1 = -\left( a_2 -
1\right)~, \ \ \ a_{-1}-1 = -\left( a_1 - 1\right)~.
\ee
We can simplify the other summations in a similar way.

Explicitly putting in the values of $a_1,a_2,a_3$ given in Eq.~(\ref{eq:a}), 
the difference equation, Eq.~(\ref{dif-eq}), becomes
\beq\label{eq:von_neumann1}
 &&e^{4\chi i} \Biggl[ T\left(a_0\lambda_1\right)
   e^{\frac{2i\omega\lambda_2\left(\nu+1\right)}{\nu
       \left(\nu+2\right)}} \cos\left(
   \frac{2\omega\lambda_2\left(\nu+1\right)}{\nu\left(\nu+2\right)}\right)
 + T\left(a_1\lambda_1\right) e^{\frac{i\omega\lambda_2}{v}}
   \cos\left( \frac{\omega\lambda_2}{\nu}\right)
\nonumber\\
&&   \ \ \ \ \ \ \ \
  + T\left(a_2\lambda_1\right) e^{\frac{i\omega\lambda_2}{\nu+2}}
   \cos\left( \frac{\omega\lambda_2}{\nu+2}\right) \Biggr] \nonumber
\\
&&=\  \sqrt{\frac{\nu}{\nu+4}} \Biggl[ T\left( a_{-1}\lambda_1\right)
  e^{\frac{-i\omega
      \lambda_2}{\nu+2}}\cos\left(\frac{\omega\lambda_2}{\nu+2}\right)
+ T\left( a_0\lambda_1\right)
  \cos\left(\frac{2\omega\lambda_2}{\nu+2}\right)
\nonumber\\
&&\ \ \ \ \ \ \ \ \ \ \ \ \ \ \ \ \ \   +T\left(
  a_1\lambda_1\right) e^{\frac{i\omega
      \lambda_2}{\nu+2}}\cos\left(\frac{\omega\lambda_2}{\nu+2}\right)
  \Biggr] \nonumber \\
&& \ \ -\left(\frac{\nu-2}{\nu+2}\right) \sqrt{\frac{\nu}{\nu+4}} \Biggl[
  T\left( a_{-2}\lambda_1\right) e^{\frac{i\omega
      \lambda_2}{\nu-2}}\cos\left(\frac{\omega\lambda_2}{\nu-2}\right)
+ T\left( a_0\lambda_1\right)
  e^{\frac{2i\omega\lambda_2}{\nu\left(\nu-2\right)} }
  \cos\left(\frac{2\omega\lambda_2}{\nu\left(\nu-2\right)}\right)
\nonumber\\
&&\ \ \ \ \ \ \ \ \  \ \ \ \ \ \ \ \ \ \ \ \ \ \ \ \ \ \ \ \ \ \ 
  + T\left( a_3\lambda_1\right) e^{\frac{i\omega
      \lambda_2}{\nu}}\cos\left(\frac{\omega\lambda_2}{\nu}\right)
  \Biggr] \nonumber \\
&& \ \ -\left(\frac{\nu-2}{\nu+2}\right) \sqrt{\frac{|\nu-4|}{\nu+4}}
e^{-4\chi i}\Biggl[ T\left( a_{-3}\lambda_1\right) e^{\frac{-i\omega
      \lambda_2}{\nu}}\cos\left(\frac{\omega\lambda_2}{\nu}\right) 
+  T\left( a_{-2}\lambda_1\right) e^{\frac{-i\omega\lambda_2}{\nu-2} }
  \cos\left(\frac{\omega\lambda_2}{\nu-2}\right) 
\nonumber\\
&&\ \ \ \ \ \ \ \ \ \ \ \ \ \ \ \ \ \ \ \ \ \ \ \ \ \ \ \ \ \ \ 
\ \ \ \ \ \ \ \   
+T\left(
  a_0\lambda_1\right) e^{\frac{-i\omega
      \lambda_2\left(\nu-1\right)}{\nu\left(\nu-2\right)}}
  \cos\left(\frac{\omega\lambda_2\left(\nu-1\right)}
{\nu\left(\nu-2\right)}\right)
  \Biggr]~. 
\eeq

Up to this point the equation is exact, however expanding in terms of
small $1/\nu$, Eq.~(\ref{eq:von_neumann1}) becomes
\beq\label{eq:diff_eqn3}
e\left(4\chi\right) \Biggl[ T\left(a_0\lambda_1\right)
  e\left(2\Lambda\right) \cos \left(2\Lambda\right) + \left(
  T\left(a_1\lambda_2\right) + T\left( a_2 \lambda_1\right) \right)
  e\left(\Lambda\right) \cos \left(\Lambda\right) \biggr] \nonumber \\
= \left\{1-\frac{2}{\nu}\right\} \Biggl[ T\left(a_0\lambda_1\right)
  \cos\left(2\Lambda\right) + T\left(a_{-1}\lambda_1\right)
  e\left(-\Lambda\right) \cos \left( - \Lambda\right) +
  T\left(a_1\lambda_1\right) e\left( \Lambda\right)
  \cos\left(\Lambda\right) \Biggr] \nonumber \\
+ \left(1-\frac{6}{\nu}\right) \Biggl[ T\left(a_{-2}\lambda_1\right)
  e\left( -\Lambda\right) \cos \left( -\Lambda\right) + T\left( a_3
  \lambda_1\right) e\left(\Lambda\right) \cos \left(\Lambda\right)
  \Biggr] \nonumber \\
- \left\{1-\frac{8}{\nu}\right\} e\left(-4\chi\right) \Biggl[ T\left(
  a_{-3} \lambda_1\right) e\left( -\Lambda\right) \cos\left(
  -\Lambda\right) + T\left( a_{-2}\lambda_1\right) e\left(
  -\Lambda\right)\cos\left( -\Lambda\right) + T\left(
  a_{0}\lambda_1\right) e\left( -\Lambda\right)\cos\left(
  -\Lambda\right) \Biggr]\nonumber\\
 +{\cal O}\left( \frac{1}{\nu^2}\right)~,
\eeq 
where we have defined the function 
\be
\label{e(x)}
e\left(x\right) = e^{ix}~,
\ee 
and the variable 
\be
\label{lambda}
\Lambda = \omega \lambda_2/\nu~.
\ee  Equation
(\ref{eq:diff_eqn3}) can be re-ordered to read
\be\label{eq:reordered} 
A T\left(a_3\lambda_1\right) = B
T\left(a_2\lambda_1\right) + C T\left(a_1\lambda_1\right) + D
T\left(a_0\lambda_1\right) + E T\left(a_{-1}\lambda_1\right) + F
T\left(a_{-2}\lambda_1\right) + G T\left(a_{-3}\lambda_1\right)~, 
\ee
where 
\beq A &=& -\left[ 1-\frac{6}{\nu}\right] e\left(\Lambda\right)
\cos \left( \Lambda\right) \nonumber \\ 
B &=& -e\left(4\chi\right)
e\left(\Lambda\right) \cos \left(\Lambda\right)\nonumber \\ 
C &=&\left[1-\frac{2}{\nu} - e\left( 4\chi\right) \right]
e\left(\Lambda\right) \cos\left(\Lambda\right) \nonumber \\ 
D &=&\left[ - e\left(4\chi\right) e\left(2\Lambda\right) +1 - \frac{2}{\nu}
\right] \cos \left(2\Lambda\right) -\left[ 1-\frac{8}{\nu}\right]
e\left( -4\chi\right) e\left(-\Lambda\right)\cos \left(
-\Lambda\right) \nonumber \\ 
E &=& \left[1-\frac{2}{\nu}\right]e\left(-\Lambda\right) 
\cos \left( -\Lambda\right) \nonumber \\ 
F&=& \left[ 1-\frac{6}{\nu} - \left\{ 1-\frac{8}{\nu}\right\}
e\left(-4\chi\right) \right] e\left( -\Lambda\right)
\cos\left(-\Lambda\right) \nonumber \\ 
G &=& -\left[ 1 - \frac{8}{\nu}\right] e\left(-4\chi\right) 
e\left(-\Lambda\right) \cos\left( -\Lambda\right)~.  
\eeq
Equation~(\ref{eq:reordered}) is equivalent to the vector equation
\be
 M_1 \overline{T}_{3} = M_2 \overline{T}_{2}~,
\ee
where we have defined the vectors
\be
 \overline{T}_i = \left[ \begin{array}{c}
 T\left(a_i \lambda_1\right) \\
 T\left(a_{i-1} \lambda_1\right) \\
 T\left(a_{i-2} \lambda_1\right) \\
 T\left(a_{i-3} \lambda_1\right) \\
 T\left(a_{i-4} \lambda_1\right) \\
 T\left(a_{i-5} \lambda_1\right) \end{array} \right]~~\mbox{for}~~i=2,3
\ee
and the matrices
\be
M_1 = \left( \begin{array}{cccccc}
A & 0 & 0 & 0 & 0 & 0 \\
0 & 1 & 0 & 0 & 0 & 0 \\
0 & 0 & 1 & 0 & 0 & 0 \\
0 & 0 & 0 & 1 & 0 & 0 \\
0 & 0 & 0 & 0 & 1 & 0 \\
0 & 0 & 0 & 0 & 0 & 1 \end{array} \right)~, \ \ \ \ \ 
M_2 = \left( \begin{array}{cccccc}
B & C & D & E & F & G \\
1 & 0 & 0 & 0 & 0 & 0 \\
0 & 1 & 0 & 0 & 0 & 0 \\
0 & 0 & 1 & 0 & 0 & 0 \\
0 & 0 & 0 & 1 & 0 & 0 \\
0 & 0 & 0 & 0 & 1 & 0 \end{array} \right)~.
\ee
Stability of this system is then given by the eigenvalues of the
matrix $\left( M_1\right)^{-1}M_2$. More particularly, if 
\be
 \max |\tilde{\lambda} | \leq 1 ~~~\forall~~\omega\ \mbox{and}\ \chi~,
\ee 
where $\tilde{\lambda}$ are the eigenvalues of the matrix
$\left(M_1\right)^{-1}M_2$, then the amplitude
$T\left(a_3\lambda_1\right)$ is less than that of previous points,
namely the difference equation is stable.

One finds, in block form, that
\be\label{eq:matrix}
 \left(M_1\right)^{-1}M_2 = \left( \begin{array}{c|c}
\tilde{A},\tilde{B},\tilde{C},\tilde{D},\tilde{E}& \tilde{F} \\
\hline
\mathbbm{1}_5 & \underline{0}_5
\end{array}\right)~,
\ee
where $\mathbbm{1}_5$ is the $5\times 5$ identity matrix,
$\underline{0}_5$ is the zero vector and 
\beq\label{eq:A-F}
\tilde{A} &=& \left[ 1 + \frac{6}{\nu}\right]e\left(4\chi\right)
 \nonumber \\ \tilde{B} &=& -\left[ 1 + \frac{6}{\nu}\right]\left[ 1 -
 \frac{2}{\nu} - e\left(4\chi\right)\right] \nonumber \\ \tilde{C}
 &=& \left[ 1 + \frac{6}{\nu}\right] \left[ e\left(4\chi\right)
 e\left(2\Lambda\right) + 1 -\frac{2}{\nu}\right]
 e\left(-\Lambda\right) \frac{\cos
   \left(2\Lambda\right)}{\cos\left(\Lambda\right)} + \left[ 1
 -\frac{2}{\nu} \right] e\left( -4\chi\right) \nonumber \\
\tilde{D} &=& -\left[ 1 + \frac{4}{\nu} \right]
e\left(-2\Lambda\right) \nonumber \\ \tilde{E} &=& -\left[ 1 +
\frac{6}{\nu} \right] \left[ 1-\frac{6}{\nu} - \left\{
1-\frac{8}{\nu}\right\} e\left(-4\chi\right) \right] e\left(
-2\Lambda\right)\nonumber \\ \tilde{F} &=& \left[
1-\frac{2}{\nu}\right] e\left( -4\chi\right) e\left(
-2\Lambda\right)~,
\eeq
with $e(x)$ as defined in Eq.~(\ref{e(x)}), previously. 
The eigenvalues of Eq.~(\ref{eq:matrix}) are found by solving the
characteristic equation
\be\label{eq:char_eq}
 \Big| \left( M_1\right)^{-1}M_2 - \tilde{\lambda}\mathbbm{1}_6 \Big| =0~,
\ee
for the eigenvalues $\tilde{\lambda}$; note that $\mathbbm{1}_6$ is the 
$6\times 6$ identity matrix. We are looking for the maximum
$|\tilde{\lambda}|$, {\it for all} $\omega$ and $\chi$. We can
immediately see that the system will not be stable, since the inverse
of $M_1$ 
only exists when $|A| \neq 0$.  The cases when $|A|=0$
correspond to
\be
\Lambda = \frac{ \left(2n-1\right) \pi}{2}~,
\ee
or, equivalently, using Eq.~(\ref{lambda}):
\be
\omega= \frac{\left(2n-1\right) \pi}{2} \frac{\nu}{\lambda_2}~,
\ee 
with $n\in \mathbb{Z}$ and these modes are explicitly {\it
  unstable}. This can be understood by noting that the amplitude
$T\left( a_3\lambda_1\right)$ is multiplied by $A$, which can be made
arbitrarily small, hence then the amplitude $T\left(
a_3\lambda_1\right)$ has to be arbitrarily large.

We can go further and consider the $0^{\rm th}$ order limit in the
$(1/\nu) \rightarrow 0$ expansion, in which the definitions given in
Eq.~(\ref{eq:A-F}) simplify to
\beq
\label{coef}
\tilde{A}^{(0)} &=& e^{4i\chi}\nonumber\\ 
\tilde{B}^{(0)} &=& -\left(1 -e^{4i\chi}\right) \nonumber \\ 
\tilde{C}^{(0)} &=& \left( e^{4i\chi +2i\Lambda} 
+ 1\right)e^{-i\Lambda} \frac{\cos 2\Lambda}{\cos \Lambda} 
+ e^{-4i\chi}\nonumber\\ 
\tilde{D}^{(0)} &=& -e^{-2i\Lambda}\nonumber \\ 
\tilde{E}^{(0)} &=& -\left(1-e^{-4i\chi}\right)e^{-2i\Lambda}\nonumber\\
\tilde{F}^{(0)} &=& e^{-4i\chi -2i\Lambda}~,
\eeq
where the superscript $^{(0)}$, reminds us that we are working to the
$0^{\rm th}$ order in the small $(1/\nu)$ expansion.  

If we further consider the modes given by $\Lambda = \pi/4$ and $\chi
=0$, then the above coefficients, Eq.~(\ref{coef}), become simply
\beq
 \tilde{A}^{(0)} &=& 1~, \ \ \ \ \tilde{B}^{(0)} = 0~,\nonumber\\
 \tilde{C}^{(0)} &=& 1~, \ \ \ \ \tilde{D}^{(0)} = i~,\nonumber\\
\tilde{E}^{(0)} &=& 0~,\ \ \ \  \tilde{F}^{(0)} = -i~.
\eeq
In this specific case, the matrix given in Eq.~(\ref{eq:matrix}) reads
\be
 M_1^{-1} M_2 = \left(\begin{array}{cccccc}
  1 & 0 & 1 & i & 0 & -i \\
  1 & 0 & 0 & 0 & 0 & 0 \\
  0 & 1 & 0 & 0 & 0 & 0 \\
  0 & 0 & 1 & 0 & 0 & 0 \\
  0 & 0 & 0 & 1 & 0 & 0 \\
  0 & 0 & 0 & 0 & 1 & 0 \end{array} \right)~,
\ee
the determinant of which is simply
\be\label{cond1}
\det \left(M_1^{-1}M_2\right) = -i~,
\ee
implying
\be\label{cond2}
-i= \Pi_{j=1}^6 \tilde{\lambda}_j;
\ee
$\tilde{\lambda}_j$ are the eigenvalues of the matrix
$M_1^{-1}M_2$. Since $| \Pi_{j=1}^6 \tilde{\lambda}_j|=1$, either $\max
\left( |\tilde{\lambda}_j|\right) > 1$, or $|\tilde{\lambda}_j| =1,
\ \forall j \ \mbox {with}\ j=1,\cdots, 6$. We can rule out the second
possibility by explicit evaluation of the characteristic equation,
Eq.~(\ref{eq:char_eq}), for this ansatz.

To be more specific, set $\tilde{\lambda}_j = \exp \left( i
\theta_j\right)$ and solve Eq.~(\ref{eq:char_eq}), subject to the
limit $(1/\nu)\rightarrow0$, for the modes $\Lambda=\pi/4$ and $\chi =0$,
to find $\theta_j$. In this case, Eq.~(\ref{eq:char_eq}) becomes
\be
\label{eq-phases}
-\left( 1-e^{i\theta_j}\right) e^{5i\theta_j} - e^{3i\theta_j} -
i\left( 1-e^{2i\theta_j}\right) =0\ \ , \forall j \ \mbox {with}\ j=1,\cdots 6~. 
\ee 
The above equation, Eq.~(\ref{eq-phases}), has only two (numeric)
solutions, which without loss of generality we denote by $\theta_1,
\theta_2$, and are approximately equal to $\theta_1 = 1.18123$ and
$\theta_2=2.30716$, for $\theta_1, \theta_2 \in \left(0,2\pi\right)$.  However,
using Eq.~(\ref{cond2}), the sum of the phases of the six eigenvalues
must satisfy
\be\label{eq:constraint} 
\sum_{j=1}^6 \theta_j = \left(2n-1\right)\pi\ \ {\rm for}\ \ n\in \mathbb{Z}~.
\ee
With only two solutions, the eigenvalues must be degenerate. Let us
suppose that $|\tilde{\lambda}_j| =1, \ \forall j \ \mbox
{with}\ j=1,\cdots, 6$, and consider $p$ eigenvalues with phase
$\theta_1$ and $q$ eigenvalues with phase $\theta_2$, where $p$
and $q$ are integers satisfying $p+q=6$. We can then look for
any combination of degeneracies ({\sl i.e.}, any values of $p$ and
$q$) that satisfy Eq.~(\ref{eq:constraint}).  Explicitly it can be
verified that there is no such solution, which implies that not all of
the eigenvalues lie on the complex unit circle and hence there must be
at least one eigenvalue with $|\tilde{\lambda}_j|>1$.

A partial proof of this result in the general case can be produced by
using a variant of the Gershgorin circle
theorem~\cite{Gerschgorin1,Gerschgorin2}. The standard theorem states that the
eigenvalues of a matrix ${\cal M}=\left(a_{ij}\right)$, lie within the
$i$ discs, $D\left(a_{ii},R\right)$ (called Gershgorin discs) in the
complex plane with centre $a_{ii}$ and radius $R=\sum_{i\neq
  j}|a_{ij}|$.  It can further be shown that if the discs are
disjoint, then there is at least one eigenvalue within each connected
region. For the case of the matrix given by Eq.~(\ref{eq:matrix}) this
implies that all of the eigenvalues lie within the discs
\be\label{eq:discs}
D\left(0,1\right)~~,~~
D\left(\tilde{A},|\tilde{B}|+|\tilde{C}|+|\tilde{D}|
+|\tilde{E}|+|\tilde{F}|\right)~.
\ee
Of the two discs, the second one is the most interesting. It is
centred at $\tilde{A}$ and one can easily check that for $(1/\nu) \neq
0$, it is beyond the unit complex circle, {\sl i.e.}, $|\tilde{A}| > 1$.
However, one can also check that the radius satisfies 
\be
|\tilde{B}| + |\tilde{C}| + |\tilde{D}| + |\tilde{E}| +
|\tilde{F}| > |\tilde{A}| - 1~,
\ee
except for small values of $\nu$. Thus, the two Gershgorin discs
intersect and we cannot say that there is an eigenvalue with
$|\lambda_j| >1$.  However, by noting that $|\tilde{C}|$ becomes
arbitrarily large for $\Lambda \rightarrow \pi/2$, one realises that
the radius of the second disc in Eq.~(\ref{eq:discs}), encompasses all
of the complex plane. This would tend to suggest that there is at
least one eigenvalue that is not constrained to have
$|\lambda_j|<1$. A variation on the proof of the standard Gershgorin
circle theorem can be used to show that this is indeed the case.

Consider the case of a matrix ${\cal M} = \left( a_{ij}\right)$ such
that $|a_{13}| \gg \sum_{j\neq 3}|a_{1j}|$.  Then the characteristic
equation is
\be
 \sum_{j=1}^6 a_{ij}x_j = \lambda x_i\ \ \forall i~,
\ee
where $x=\left(x_i\right)$ is the eigenvector of ${\cal M}$ and
$\lambda$ is the corresponding eigenvalue. Expanding this sum as
\be
 a_{i3} x_3 + \sum_{j\neq 3} a_{ij} x_j = \lambda x_i~,
\ee
gives
\be
 \left| \lambda - a_{i3}\frac{x_3}{x_i} \right| = \left| \sum_{j\neq
   3} a_{ij} \frac{x_j}{x_i} \right|~,
\ee
which is valid, provided $x_i \neq 0$. If we take $x_i$ to be 
\be
x_i = \max ( x_j)\ \ \mbox{for}\ \ j\neq 3~,
\ee 
we have
\be
 \left| \lambda - a_{i3} \frac{x_3}{x_i} \right| \leq \sum_{j\neq 3} 
|a_{ij}|~.
\ee
Thus, the eigenvalue $\lambda$ is within a disc, centred at the point
$a_{i3}x_3/x_i$ with radius given by the sum of the magnitudes of the
elements along the $i^{\rm th}$ row of ${\cal M}$, {\it excluding} the
third element. In particular, if $|a_{13}|\nearrow \infty$, then for
$x_3/x_1 >0$, the centre of the disc tends to infinity. Provided the
sum $\sum_{j\neq 3}| a_{1j}|$ remains finite, the
eigenvalue $\lambda$ will lie within a disc that is entirely {\it
  outside} the complex unit circle and hence $|\lambda |>1$. This is
precisely the situation we have for ${\cal M} = M_1^{-1}M_2$, in the
case of $\Lambda\rightarrow \pi/2$.

The final element that is required for this proof is that $x_3/x_1>0$
or, more precisely, that $a_{13}x_3/x_1 \gg \sum_{j\neq 3} |a_{1j}|$,
given that $|a_{13}|\gg \sum_{j\neq 3} |a_{1j}|$ . In the
particular case of the matrix given by Eq.~(\ref{eq:matrix}), we can
evaluate the simultaneous equations implied by the characteristic
equation, Eq.~(\ref{eq:char_eq}), to find
\beq
 \tilde{C}x_3 &\approx& \lambda x_1~,\ \ \ x_1 = \lambda x_2~,\ \ \ 
x_2 = \lambda x_3~, \nonumber \\
x_3 &=& \lambda x_4~,\ \ \ x_4 = \lambda x_5~,\ \ \ x_5 = \lambda x_6~,
\eeq
where we have used the approximation that $\tilde{C}$ dominates the
terms in $\sum_{i}|a_{1i}|$. This gives
\be
 \left | a_{13}\frac{x_3}{x_1}\right| \approx \left| \tilde{C}\right|^{1/3}~.
\ee
Thus, provided that $\left| \tilde{C}\right|^{1/3} \gg \sum_{j\neq 3}
|a_{1j}|$, the proof is valid and we have
$\max\left(|\lambda_i|\right)>1$. Note that this condition is
certainly met as $\Lambda \rightarrow \pi/2$, since $\tilde{C}$
diverges, whilst $\sum_i \left|a_{1i}\right|$ remains finite. This is
essentially the result we preempted in the comments following
Eq.~(\ref{eq:char_eq}), however here we have explicitly extended it to
the case of $\tilde{C}$ large, but not infinite ({\sl i.e.}, the case when
$M_1$ is invertible, but $A$ is large).

\section{Conclusions}
The aim of this paper is to study the stability of the Hamiltonian
constraint equation valid for anisotropic Bianchi~I LQC.  Performing a
von Neumann stability analysis, we have shown that if the difference
equation admits solutions with amplitudes that grow locally, then it
is not locally stable. On the one hand, this result certainly
questions the validity of the quantisation, since any semi-classical
solutions would quickly become dominated by the expanding spurious
ones.  On the other hand however, the presence of such an instability
may not be, necessarily, a problem, since it might be that the
unstable trajectories are explicitly removed by the physical inner
product.

More precisely, the difference equation, given by
Eq.~(\ref{eq:diff_eqn}), is unconditionally unstable.  By this we mean
that there is no region of $\left(\lambda_1,\lambda_2,\nu\right)$ in
which the difference equation, Eq.~(\ref{eq:diff_eqn}), is stable. It
is worth noting however, that in Eq.~(\ref{eq:reordered}) we choose to
re-order the difference equation in such a way that it produces a
single amplitude ($T\left(a_3\lambda_1\right)$ in
Eq.~(\ref{eq:reordered})), given the other $23$ amplitudes. This is
clearly an {\it explicit} implementation of the equation. It is also
possible that this difference equation could be implemented via an
{\it implicit} scheme, {\sl i.e.}, that the equation could be
re-ordered to give (say) two amplitudes, given the values of the other
$23$ or $22$ amplitudes. In order for the system to give solutions,
one would then have to implement consistency relations between the
calculated amplitudes at different iterations. There are, of course,
many ways that such an implicit implementation of the difference
equation could be under taken and they could, in principle, have
different stability properties.

We have demonstrated the presence of an instability in the {\it
  explicit} implementation of the difference equation,
Eq.~(\ref{eq:diff_eqn}), in several ways: we have first shown that for
a particular set of critical modes, $\Lambda=(2n-1)\pi/2$, the system
is unstable. We have then showed that in the large $\nu$ limit, the
system is again unstable for the modes $\Lambda = \pi/4$ and $\chi =
0$. Finally, we have formally showed that the system is unstable for a
general $\nu$, for modes that approach the critical value. This was
done via a version of the Gershgorin circle theorem, which have
explicitly demonstrated the instability, even for modes approaching
(but not reaching) the critical value.

\vskip.05truecm 
\acknowledgements 
It is a pleasure to thank Martin Bojowald for discussions. The work of
M.S. is partially supported by the European Union through the Marie
Curie Research and Training Network \emph{UniverseNet}
(MRTN-CT-2006-035863).

\end{document}